\documentclass[useAMS,usenatbib,usegraphicx]{mn2e}

\usepackage{amssymb}

\newcommand{\etal}{{et al.}}

\title[Quantitative measure of evolution of bright cluster galaxies]{Quantitative measure of evolution of bright cluster galaxies at moderate redshifts}

\author[Vinu Vikram et al.]{Vinu Vikram,$^{1,3}$\thanks{E-mail: vvinuv@iucaa.ernet.in} 
Yogesh Wadadekar,$^{2}$\thanks{E-mail: yogesh@ncra.tifr.res.in} 
Ajit K. Kembhavi$^{3}$\thanks{E-mail: akk@iucaa.ernet.in} 
\newauthor
and G. V. Vijayagovindan$^{1}$\thanks{Deceased}\\
$^{1}$School of Pure and Applied Physics, Mahatma Gandhi University, Kottayam, Kerala, India\\
$^{2}$National Centre for Radio Astrophysics, Post Bag 3, Ganeshkhind, Pune 411007\\
$^{3}$IUCAA, Post Bag 4, Ganeshkhind, Pune 411007}

\begin{document}
\date{Accepted 2009 October 31.  Received 2009 October 31; in original form 2009 September 27}
%\pagerange{\pageref{firstpage}--\pageref{lastpage}} \pubyear{2009}
\maketitle
\label{firstpage}

\begin{abstract}
Using archival data from the Hubble Space Telescope, we study the
quantitative morphological evolution of spectroscopically confirmed
bright galaxies in the core regions of nine clusters ranging in
redshift from $z = 0.31$ to $z = 0.84$.  We use morphological
parameters derived from two dimensional bulge-disk decomposition to
study the evolution.  We find an increase in the mean bulge-to-total
luminosity ratio $B/T$ as the Universe evolves. We also find a
corresponding increase in the fraction of early type galaxies and in
the mean S\'ersic index. We discuss these results and their
implications to physical mechanisms for evolution of galaxy
morphology.
\end{abstract}


\section{Introduction}
Clusters serve as laboratories for investigating the dependence of
galaxy morphology on the density of the environment.  \citet{dre80}
found that the fraction of early type galaxies increases with local
density of galaxies.  He also found that around 80\% of galaxies in
nearby clusters are of the early type.  In recent years, deep, high
resolution imaging by the \textit{Hubble Space Telescope (HST)} has
helped to extend the study of galaxy morphology as a function of the
environment to $z \sim 1$.  The observed fraction of different
morphological types can be explained in terms of both 'nature' and
'nurture' scenarios.  The former argues that a galaxy's morphological
type is determined by initial conditions at formation
\citep[e.g. ][]{egg62}, while the latter depends on the influence of the
environment and of secular evolution for determining the final
morphological type \citep[e.g. ][]{too77}. Numerical simulations play an
important role in investigating the relative importance of the two
scenarios under different physical conditions.

Following Dressler's pioneering work, many observational studies have
been done to measure and understand the morphology density relation
(MDR), and the dependence of morphological type on distance of the
galaxy from the cluster centre
\citep{dre97,got03,smi05,pos05,hol07,tre03,cap07}.  \citet{smi05}
found that the early type fraction is constant in low density
environments over the last 10 Gyr, but there is significant evolution
in this fraction in higher density regions. According to
\citet{smi05}, this suggests that most of the ellipticals in clusters
formed at high redshift, and the increase in the fraction of early
type galaxies is because of the physical processes in dense regions
which transform disk galaxies with ongoing star formation to early
types. Dissipationless merging of cluster galaxies may also be
responsible for this increase.

There is increasing evidence to show that massive ellipticals formed
by dissipationless (dry) merger of two or more systems.  Such
ellipicals must have formed at later times than their low luminosity
counterparts \citep{luc06}.  \citet{dok05} analysed tidal debris of
elliptical galaxies and concluded that $\sim70\%$ of the bulge
dominated galaxies have experienced a merger.  The analysis of nearby
bulge dominated galaxies has shown that the gas to stellar mass ratio
is very small and these mergers are mostly 'dry'.  Using a 
semi-analytic model for galaxy formation, \citet{naa06} found that
both the photometric and kinematic properties of massive elliptical
galaxies are in agreement with the scenario where massive elliptical
galaxies are produced by mergers of lower mass ellipticals.  They
suggested that the merger of two spiral galaxies alone cannot
reproduce the observed properties, and that the large remnant mass ($>
6\times 10^{11} M_\odot$) implies that they must have undergone
elliptical - elliptical mergers.  They found that this process is
independent of the environment and redshift, which means that dry
mergers can occur at low redshift as well.  By analysing spirals from
cluster and group environments at intermediate redshift ($z\sim 0.5$),
\citet{mor07} found that the Tully-Fisher relation shows larger
scatter for cluster spirals than for those in the field.  They also
found that the central surface mass density of spirals in clusters is
small beyond the cluster virial radius, and argued that these
observations provide evidence for merger/harrasment.

Mergers are not common in clusters, as the velocity dispersion of
virialized clusters is large.  So if ellipticals in clusters have
formed by mergers, that most likely happened during the early stage of
cluster collapse \citep{roo82}. There is some observational evidence
to support this idea; \citet{dok99} showed that there is a large
fraction of ongoing mergers in a cluster at $z = 0.83$.  The
observation of the unvirialized cluster $\textrm{RX J}0848+4453$ at
$z=1.27$ revealed many ongoing dissipationless mergers of galaxies
\citep{dok01}.  These observations go against the view of monolithic
collapse where all the ellipticals formed at the same time, at very
high redshift.

In this letter we report on the evolution of galaxies in the core
region of clusters, measured using bulge-disk decomposition.  We show
that in the case of brightest cluster galaxies, the fraction of
galaxies with $B/T>0.4$ and $n>2.5$ evolved significantly over the
redshift range 0.31 to 0.83.  Throughout the paper we use the standard
concordance cosmology with $\Omega_\Lambda = 0.73$, $\Omega_m = 0.27$
and $H_0 = 71$ km s$^{-1}$ Mpc$^{-1}$.

\section{Cluster sample and Decomposition Technique}

We have exclusively used archival Hubble Space Telescope (HST) data in
this work. All of the observations used by us were obtained with
either the ACS or WFPC2 cameras on-board the HST. We constructed our
sample of clusters by an extensive literature survey of HST
observations of moderate redshift clusters.  The clusters we study
span the redshift range 0.31 to 0.837. The clusters were selected such
that they each had at least 15 spectroscopically confirmed cluster
members listed in the literature
\citep{cou98,sma97,dre99,wil99,hal04,dem05}.  We restricted our study
only to those clusters whose second brightest member galaxy has an
absolute magnitude between -25.0 and -27 in rest-frame B-band. Here
the magnitudes are corrected for the cosmological surface brightness
dimming. This criterion allows us to restrict ourselves to clusters of
roughly comparable luminosity.  Further, we only included clusters
with imaging data around the rest-frame $B$ filter i.e.  in filters
where the central wavelength of the filter corresponded to a rest
frame wavelength in the range 350-550 nm. This has the advantage that
the k-correction for transformation from $R$ or $I$ filters in the
observer frame, to the $B$ filter in the rest frame, has only a weak
dependence on galaxy spectral type in the redshift range $0.3 \lesssim
z \lesssim 0.8$ \citep{boh07}. Incompleteness of the spectroscopy is a
potential problem; since our data are drawn from a heterogeneous set
of observations, it is not possible to correct for incompleteness in a
completely consistent way. However, for the five clusters in our
sample which are part of the ESO Distant Clusters Survey (EDisCS),
\citet{hal04} have shown that incompleteness does not introduce a
significant bias. The level of incompleteness is also relatively small
because they (like us) restrict themselves to the brighter cluster
members. Nine clusters satisfied our selection conditions at the time
of commencing this study; the basic data for these clusters are given
in Table \ref{clustab}.

The images we downloaded from the HST data archive were processed in a
standard way using the On-the-Fly Reprocessing (OTFR) pipeline at
STScI. The processed images were dark and bias subtracted and
flat-fielded by the OTFR pipeline.  We combined them using the
Multidrizzle package \citep{koe02} to flag and remove cosmic rays and
to correct for geometric distortion and produced a single coadded
image. For some clusters, multiple disjoint pointings have been used;
in such cases we obtained multiple multidrizzled images. The
correlated pixel noise introduced by the drizzle process, was
corrected for, using the prescription of \citet{cas00}.

We computed the center of each cluster as the centroid of the
brightest cluster galaxy. We then selected all galaxies with a
spectroscopic redshift confirming their cluster membership and located
within a 1 Mpc projected distance from the cluster center.  Our study
is therefore restricted to galaxies lying (mostly) within the core
region of the clusters.  Note that for three clusters -- AC 114, 
CL 0303+17, and 3C 295 -- the HST imaging is not complete for the 1 Mpc
projected distance.

Galaxies are known to undergo luminosity evolution, to account for which we
adopted the following simple scheme: $M_V(z) = M_V(z=0) - 0.8 z$, where $z$ is the redshift of the object and
$M_V(z=0) = -19.5$ \citep{pos05}.  The magnitude cutoff in the observed HST filter was then calculated using

\begin{equation}
 m_{lim} = M_V(z) + DM - (M_V - M_{HST}) + k_{HST}
\end{equation}
where DM is the distance modulus, $M_V - M_{HST}$ is the rest-frame
color between  $V$ and the observed HST filter and $k_{HST}$ is the
k-correction in the observed HST filter. The k-correction for
each cluster was computed using the elliptical SED provided by \citet{pog97}.
All magnitudes are in the Vega system.

With all the above constraints, we obtained a sample of 379 galaxies
in nine clusters.  We used the GALFIT \citep{pen02} program for 2D
bulge disk decomposition of the galaxy images.  To fit the galaxies,
we have developed an automated pipeline (Vinu \etal \ 2010, in
preparation) which completely automates the fitting procedure and
organizes the results into a database, supplemented by a number of
diagnostic plots to detect anomalies.  The basic steps in performing
the 2D decomposition are: (1) make a cut-out image for each galaxy,
including neighbouring galaxies.  (2) estimate starting values of
fitting parameters using the parameters measured by Sextractor (3)
make masks to exclude stars, faint galaxies and other artifacts and
(4) Run GALFIT. Bright neighboring galaxies were fitted simultaneously
with the target galaxy.  We modeled each galaxy as a linear sum of a
S\'ersic (for the bulge) and an exponential (for the disk)
component. The centers, position angles, ellipticities and central
surface brightnesses of bulge and disk were fitted simultaneously. The
S\'ersic index was left unconstrained. The estimation of galaxy
parameters, particularly the S\'ersic index, may be systematically affected
 if the sky or PSF are incorrectly estimated. To
test for the effect of using an incorrect PSF, we ran the decomposition for
each galaxy using two stellar PSFs, one constructed using the nearest
star and the other using the second nearest star and compared the
fitted values of $B/T$ and $n$ for every galaxy. To test for incorrect
sky, we ran the decomposition in two modes: one in which the sky was
left as a free parameter and one in which it was fixed to the local
sky value as determined by Sextractor. In both tests, $\sim80$\%
of galaxies showed random changes in the extracted parameters at $< 10$\% level,
with no obvious systematics. Our simple tests are consistent with the extensive simulations of \citet{hau07} which showed that GALFIT estimates parameters accurately for HST data,
even at relatively shallow depths. As an additional precaution, we
examined the fit results, by eye, for every galaxy. The fit diagnostic
plots were used to evaluate the quality of the fit. We used the
reduced $\chi^2$ given by GALFIT to identify galaxies with large
residuals. A large residual in the central part of the galaxy may be
caused by improper estimation of the point spread function
(PSF). Also, simultaneous fitting with neighbor galaxies needs special
care as the number of free parameters is significantly larger.  For
galaxies that show a large residual, we took these caveats into
consideration and refit the galaxies until the residual became
small. We then assigned a quality factor for each galaxy based on the
magnitude of the residual and the deviation of fitted position angle
from the galaxy position angle estimated by eye. If a fit was below a
threshold quality, we excluded that galaxy from further analysis. We
found that most of the galaxies that failed the quality check, either
have peculiar morphology or strong spiral arms. We successfully fit
337 out of 379 galaxies in our sample. Unless otherwise stated, all
further discussion in this letter only applies to these 337 galaxies.
The cluster-wise breakup of galaxies with a good fit is given in Table
\ref{clustab}.

\begin{table*}
 \centering
 \begin{minipage}{150mm}
  \caption{Summary data for sample clusters} 
  \label{clustab}
  \begin{tabular}{@{}llllllllll@{}}
  \hline
Cluster & RA & Dec & z & Camera & Filter & $m_{lim}$ & $N_{Tot}$ & $N_{Fit}$ &  Reference\\
\hline
AC 114            & 22 58 48.4 & -34 48 60 & 0.31  & WFPC2 & F702W & 20.77 & 72 & 68 & \citet{cou98}\\
CL 0303+17        & 03 06 15.9 & +17 19 17 & 0.42  & WFPC2 & F702W & 21.64 & 28 & 26 & \citet{sma97}\\
3C 295            & 14 11 19.5 & +52 12 21 & 0.46  & WFPC2 & F702W & 21.93 & 37 & 32 & \citet{sma97}\\
CL 1232-1250      & 12 32 30.3 & -12 50 36 & 0.54  & ACS   & F814W & 21.66 & 46 & 41 & \citet{whi05}\\
CL 1054-1146      & 10 54 24.4 & -11 46 19 & 0.697 & ACS   & F814W & 22.44 & 30 & 25 & \citet{whi05}\\
CL 1040-1155      & 10 40 40.3 & -11 56 04 & 0.704 & ACS   & F814W & 22.47 & 25 & 19 & \citet{whi05}\\
CL 1054-1245      & 10 54 43.5 & -12 45 51 & 0.75  & ACS   & F814W & 22.68 & 29 & 28 & \citet{whi05}\\
CL 1216-1201      & 12 16 45.3 & -12 01 17 & 0.794 & ACS   & F814W & 22.89 & 50 & 43 & \citet{whi05}\\
RX J0152.7-1357   & 01 52 27.4 & -13 55 01 & 0.837 & ACS   & F775W & 23.50 & 62 & 55 & \citet{bla06}\\
\hline
\end{tabular}

$m_{lim}$ : Faint magnitude cutoff in the observed HST filter,  $N_{Tot}$ : Total number of galaxies, $N_{Fit}$ : Number of galaxies with good fit
\end{minipage}
\end{table*}

\section{Results}
\subsection{Evolution of mean bulge-to-total luminosity ratio ($\langle B/T \rangle$) and S\'ersic index $n$}

\begin{figure}
 \centering
 \includegraphics[scale=0.35]{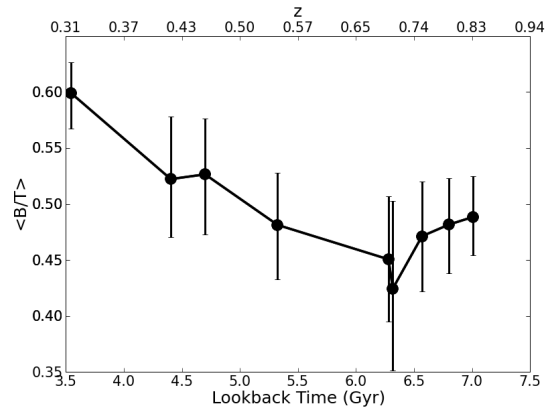}
 \caption{Evolution of mean value of the bulge-to-total luminosity ratio $B/T$}
 \label{avg-BT}
 \end{figure}

In Table \ref{results} we list mean and median values of a few
parameters of interest for all the nine clusters. Note that the
bulge-to-total luminosity ratio is computed using the parameters of
the best fit model. We find the $\langle B/T \rangle$ of cluster
galaxies in the central 1 Mpc of the clusters changes from
0.59$^{+0.03}_{-0.02}$ at redshift z = 0.31 to 0.48 $\pm$ 0.03 at z =
0.837 (Figure \ref{avg-BT}).  Errors were measured using the bootstrap
resampling method \citep{efr93}, in this and subsequent figures. The
increase in $\langle B/T \rangle$ ratio has been found qualitatively
(i.e. using visual morphological classification) by previous studies
\citep{dre97,fas00,smi05}; the present work obtains the result
quantitatively using bulge-disk decomposition. It must be noted that,
if our sample is somewhat biased towards luminous red galaxies at
high-z, and if such galaxies are early type even at those redshifts,
then our estimated $\langle B/T\rangle$ value represents an upper
limit at $z \sim 0.8$. In addition, incompleteness will also tend to
make our estimate of $\langle B/T\rangle$ too high because at fainter
magnitudes we would preferentially miss low $B/T$ late-type galaxies
\citep{des07}.  This leads to our estimated $\langle B/T\rangle$ to be
the upper limit for these clusters. So the selection bias, if any,
will lead to an apparent weaker evolution. So our estimation of
evolution of $\langle B/T\rangle$ may be an underestimate.

\begin{figure}
 \centering
 \includegraphics[scale=0.35]{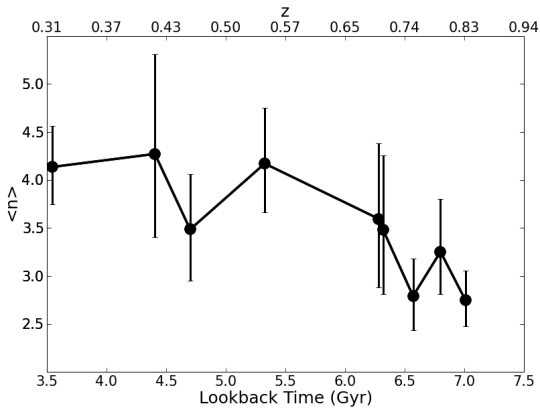}
 \caption{Evolution of mean value of S\'ersic index}
 \label{avg-n}
\end{figure}

We also find that the mean value of S\'ersic index decreases with
lookback time over this redshift range.  The value changes from
4.13$^{+0.43}_{-0.39}$ to 2.74$^{+0.28}_{-0.30}$ from $z = 0.31$ to $z
= 0.84$.  Figure \ref{avg-n} shows the evolution of S\'ersic index
against lookback time.

\begin{table}
  \caption{Mean and median parameters obtained through bulge-disk decomposition}
  \label{results}
  \begin{tabular}{@{}lllll}
  \hline
Cluster & $\langle B/T\rangle$ & $\langle n\rangle$ & \~{n} & $f_b$\\
\hline
AC 114 & $0.59^{+0.03}_{-0.02}$ & $4.13^{+0.43}_{-0.39}$ & $3.47^{+0.27}_{-0.30}$ & $0.55^{+0.03}_{-0.04}$\\
CL 0303 & $0.52^{+0.05}_{-0.05}$ & $4.24^{+1.02}_{-0.79}$ & $3.31^{+0.20}_{-0.83}$ & $0.46^{+0.05}_{-0.07}$\\
3C 295 & $0.52^{+0.04}_{-0.05}$ & $3.47^{+0.59}_{-0.56}$ & $3.17^{+0.72}_{-0.78}$ & $0.43^{+0.03}_{-0.04}$\\
CL 1232-1250 & $0.48^{+0.04}_{-0.04}$ & $4.15^{+0.56}_{-0.52}$ & $3.30^{+0.41}_{-0.52}$ & $0.47^{+0.04}_{-0.05}$\\
CL 1054-1146 & $0.45^{+0.05}_{-0.05}$ & $3.60^{+0.76}_{-0.65}$ & $2.41^{+1.02}_{-0.43}$ & $0.26^{+0.05}_{-0.06}$\\
CL 1040-1155 & $0.42^{+0.08}_{-0.07}$ & $3.50^{+0.68}_{-0.68}$ & $3.25^{+0.71}_{-0.56}$ & $0.28^{+0.08}_{-0.08}$\\
CL 1054-1245 & $0.47^{+0.05}_{-0.05}$ & $2.80^{+0.38}_{-0.41}$ & $2.55^{+0.56}_{-0.55}$ & $0.37^{+0.05}_{-0.06}$\\
CL 1216-1201 & $0.48^{+0.03}_{-0.04}$ & $3.25^{+0.54}_{-0.44}$ & $2.57^{+0.40}_{-0.51}$ & $0.32^{+0.04}_{-0.04}$\\
RX J0152.7-1357 & $0.48^{+0.03}_{-0.03}$ & $2.74^{+0.28}_{-0.30}$ & $2.49^{+0.43}_{-0.28}$ & $0.40^{+0.02}_{-0.02}$\\
\hline
\end{tabular}

\medskip
$\langle B/T\rangle$ : mean value of $B/T$ bulge-to-total luminosity ratio, $\langle n\rangle$ : mean value of Se\'rsic index, \~{n} : median value of Se\'rsic index, $f_b$ : fraction of bulge-like galaxies (see text for definition)
\end{table}

\subsection{Evolution of bulge-like galaxy fraction}

\begin{figure}
 \centering
 \includegraphics[scale=0.35]{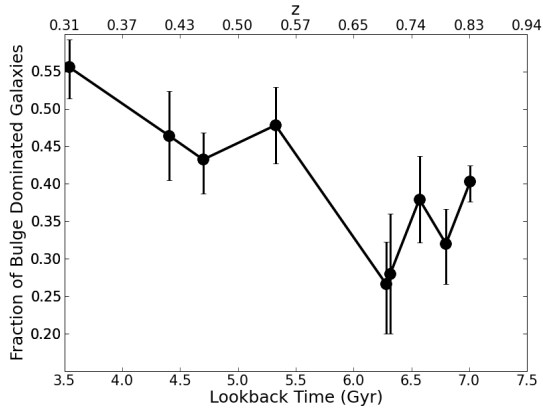}
 \caption{Evolution of fraction of bulge-dominated galaxies}
 \label{evo-early}
\end{figure}

Since the $\langle B/T\rangle$ decreases significantly with lookback
time, one expects to see a simultaneous decrease in the fraction of
early-type galaxies with a bulge-like morphology.  We classify a
galaxy as bulge-like if its $B/T \geq 0.4$ and $n > 2.5$.  The second
condition is required to exclude galaxies with a very strong disk
which can, on occasion, be incorrectly modeled as a bulge with
$n\sim1$ (an exponential), with a correspondingly high $B/T$
luminosity ratio.  With this definition, we are able to compute a
bulge-like galaxy fraction for each cluster. In figure \ref{evo-early}
we plot this fraction against lookback time.  Note that the fraction
is normalized by the total number of galaxies in the cluster, not the
number of galaxies with a successful fit.  The galaxies that are
poorly fit are dominated by galaxies of irregular morphology.  We see
a near monotonic decrease with lookback time in the bulge dominated
fraction of galaxies.  We find 40.0 $^{+2}_{-2}$ \% of galaxies at
redshift z = 0.837 are bulge-like.  This increases to 55 $^{+3}_{-4}$
\% within $\sim$ 3.5 Gyr.

\subsection{Discussion}

In the last decade, several studies have reported evolution of the 
morphological content of galaxy clusters by a visual study of galaxy
morphology \citep{sma97,cou98,fas00,des07}.  Other studies focused on
the evolution of the morphology-density relation (MDR), where
morphology changes were studied as a function of redshift and galaxy
density \citep{dre97,tre03,pos05,smi05}.

In this work, we have taken the first approach; however, we have used
{\it quantitative} measures of galaxy morphology rather than a
qualitative morphological classification by eye.  We decompose the
galaxy light into bulge and disk components and study the evolution of
morphology of galaxies in clusters.  We find that the bulge component
of the {\it bright} galaxies in clusters is, on average, becoming
stronger as the Universe evolves.  The fraction of bulge-like
galaxies, defined as having $B/T > 0.4$ and $n > 2.5$ also increases
from 40\% to 55\%.
There is also a significant increase in the mean value of the S\'ersic
index as the Universe ages.

Our results on the evolution of morphological fraction are consistent
with previous work on the subject \citep{dre97,pos05,smi05}.  However,
\citet{des07}, using ACS observations of galaxy clusters with $0.5 < z
< 0.8$ found no evolution of morphological fraction.  At first glance,
this seems to contradict our results.  However, it must be noted that
in our study, we are only including bright, spectroscopically
confirmed cluster galaxies.  If fainter galaxies are included, as was
done in the sample of \citet{des07}, the $\sim10$\% change we see in
the morphological fraction may easily be washed out.  This explanation
also agrees with the results of \citet{hol07} who showed that there is
evolution in the early type fraction with redshift if a luminosity limited sample is used. 

\citet{pos05,des07,pog09} found that the E+S0 fraction correlates with the velocity dispersion of the cluster. The fraction is large for clusters with high dispersion. This effect would be consistent with the evolution we see, provided velocity dispersion systematically decreases with redshift. We see no systematic dependence of velocity dispersion with redshift except that the two highest redshift clusters in our sample have high velocity dispersions ($> 1000 $ km/s). This high value, which implies higher  $\langle B/T\rangle$ is consistent with the mild increase of $\langle B/T\rangle$ and the fraction of bulge dominated galaxies for the two high $z$ clusters. But the dominating effect seems to be morphological evolution.

The $\langle B/T\rangle$ of a cluster may increase either due to an increase in
the strength of the bulge component of galaxies or by the fading of
disk component.  The increase in average bulge strength may be caused
by merging which tends to produce a elliptical like morphology
\citep{alb82,bar92,her92,bou05}.  On the other hand, the fading
of the disk could be a by-product of the morphological transformation
of galaxies.  In clusters, a variety of mechanisms such as galaxy
harassment \citep{moo96,moo98,moo99}, minor mergers and ram pressure
stripping \citep{gun72,aba99} may contribute to the disk fading.

The increase in the mean S\'ersic index $\langle n \rangle$ with cosmic time seems
to indicate that mergers play a role \citep{sca03}.  Merger events are common at
intermediate redshift \citep{dre94}.  The increase in the fraction of
galaxies with high $B/T$ and S\'ersic index is possible if galaxies
gradually evolve into a phase where the spheroidal component
increasingly dominates. The end point of such evolution is an elliptical
galaxy. 

It has recently been suggested that the bright S0 population has likely
formed through monolithic collapse or major mergers
\citep{bar07,bar09}.  Numerical simulations have also shown that
dissipative merger of two unequal mass disk galaxies
\citep{bek98,bek05} can produce lenticulars.  Recent observations of
low redshift clusters suggest that a majority of the infall population
is merging or interacting \citep{mos06}.  Coupled with the constraints
from \citet{car86}, it is becoming increasingly clear that formation
of spheroidal (mostly lenticular) galaxies through mergers is the
dominant mechanism behind systematic changes in galaxy morphology.  

It must be noted that the trends we see are weak and statistical in
nature; they are only visible when averaged over a large number of
galaxies in a large number of clusters over a wide range of redshift.
The detailed physics operating in each cluster, doubtless modifies the
morphological evolution of galaxies in that cluster. Nevertheless, the
fact that we see trends indicates that they are real and strong enough
not to be drowned by the different physical conditions and processes
operating in individual clusters. Using the large database of bulge/disk
decomposition results we have obtained, we are attempting to
disentangle cluster specific effects from cosmological ones.

\section*{Acknowledgments}

 Vinu Vikram acknowledges financial support from the Council of Scientific and Industrial Research (CSIR). We thank the referee for insightful comments and suggestions that greatly improved this paper.

\bsp

\label{lastpage}
\end{document}